\documentclass[12pt]{article}

\usepackage{graphicx}
\usepackage{setspace}
\usepackage[numbers]{natbib}
\usepackage{amsmath, amsthm, amsfonts, amssymb}

\usepackage{authblk}
\usepackage{mathrsfs}
\usepackage{enumerate}
\usepackage{xcolor}
\usepackage{pslatex}

\DeclareMathOperator{\erfc}{erfc}

\renewcommand{\ge}{\geqslant}
\renewcommand{\le}{\leqslant}

\title{Distribution of genotype network sizes in sequence-to-structure 
genotype-phenotype maps}
\author[1,2]{Susanna Manrubia}
\author[1,3,4,5]{Jos\'e A. Cuesta}
\affil[1]{Grupo Interdisciplinar de Sistemas Complejos (GISC), Madrid}
\affil[2]{Dept. de Biolog\'{\i}a de Sistemas, Centro Nacional de
Biotecnolog\'{\i}a (CSIC), Madrid, Spain}
\affil[3]{Dept. de Matem\'aticas, Universidad Carlos III de Madrid, Legan\'es,
Madrid, Spain}
\affil[4]{Instituto de Biocomputaci\'on y F\'\i sica de Sistemas Complejos
(BIFI)\\ Universidad de Zaragoza, Spain}
\affil[5]{UC3M-BS Institute of Financial Big Data (IFiBiD), Universidad Carlos
III de Madrid, Getafe, Madrid, Spain}

\date{}

\begin{document}

\maketitle


\begin{abstract}
An essential quantity to ensure evolvability of populations is the navigability
of the genotype space. Navigability, understood as the ease with which alternative
phenotypes are reached, relies on the existence of sufficiently large and mutually 
attainable genotype networks. The size of genotype networks (e.g. the number of RNA 
sequences folding into a particular secondary structure, or the number of DNA 
sequences coding for the same protein structure) is astronomically large in all 
functional molecules investigated: an exhaustive experimental or computational 
study of all RNA folds or all protein structures becomes impossible even for 
moderately long sequences. Here, we analytically derive the distribution of 
genotype network sizes for a hierarchy of models which successively incorporate 
features of increasingly realistic sequence-to-structure genotype-phenotype maps. 
The main feature of these models relies on the characterization of each phenotype 
through a prototypical sequence whose sites admit a variable fraction of letters of 
the alphabet. Our models interpolate between two limit distributions: a power-law 
distribution, when the ordering of sites in the prototypical sequence is strongly 
constrained, and a lognormal distribution, as suggested for RNA, when different 
orderings of the same set of sites yield different phenotypes. Our main result is 
the qualitative and quantitative identification of those features of 
sequence-to-structure maps that lead to different distributions of genotype network 
sizes. 
\end{abstract}


\section{Introduction}

How genotypes map into phenotypes counts amongst the most essential questions to 
understand how evolutionary innovations might come about and how evolutionarily 
stable strategies are fixed in populations. With some of its features seemingly 
dependent on the system studied and on the description level considered, the 
genotype-phenotype (GP) map appears far from trivial. Many studies have addressed 
the effect of mutations on phenotype: point 
mutations~\cite{lipman:1991,holzgrafe:2011,manrubia:2013}, genome fragment 
deletion~\cite{weirauch:2010}, duplication or inversions, or the knockout of 
specific genes~\cite{baba:2006} ---among others--- may or may not have an effect at 
the molecular, metabolic, regulatory, or organismal level~\cite{rutherford:2000}. 
Also, the ability of genotypes to yield more than one phenotype is a main resource 
of molecular adaptation~\cite{bloom:2007b,piatigorsky:2007}. The probability of 
expressing different phenotypes or of experiencing mutations that modify the 
current phenotype depends on the structure of the GP map, which eventually 
determines how the space of function is explored, and what are the chances that a 
population survives or innovates in the face of endogenous or exogenous 
changes~\cite{draghi:2010,wagner:2011b,schaper:2014,payne:2014,greenbury:2016,catalan:2017}.

Most models are restricted to the many-to-one realisation of the GP map, and thus
assume that adaptation is dominated by mutations. There is a plethora of 
different model systems studied under this assumption. Despite seemingly relevant 
underlying molecular differences, those models present a remarkable number of 
common properties. Exhaustive research on the GP map was pioneered by studies of 
RNA sequence-to-secondary-structure mappings. Most topological properties 
identified in RNA spaces are shared by other simple systems, such as the existence 
of huge genotype networks, the increase in phenotype robustness with the size of 
the latter, and a very skewed distribution of network sizes. The set of genotypes 
that yield the same phenotype typically forms a network, since those genotypes are 
pairwise connected through mutations. Sufficiently large genotype networks so
defined were postulated as a condition for the navigability of sequence space long 
ago~\cite{maynard-smith:1970}. Subsequent studies have shown that such large 
networks do exist, and that the difference in sequence between genotypes in those 
networks can be as large as the difference between two random 
sequences~\cite{rost:1997,ciliberti:2007,matias-rodrigues:2009,holzgrafe:2011}.
Phenotype robustness refers to the average effect of point mutations in the 
genotypes of a specific genotype network. It has been shown to grow logarithmically 
with the size of the phenotype in RNA~\cite{aguirre:2011}, in a self-assembly model 
of protein quaternary structure~\cite{greenbury:2014} and in simple models for 
protein folding~\cite{greenbury:2016}. The existence of qualitative and 
quantitative statistical properties of the GP maps shared by apparently dissimilar 
systems suggests that they might arise from basic universal 
features~\cite{wagner:2011,greenbury:2016}. Though genotype networks are not always 
fully connected, they do traverse the whole space of genotypes for sufficiently 
abundant phenotypes, thus ensuring high 
navigability~\cite{gravner:2007,mcleish:2015}. Even in cases where genotype 
networks are fragmented, those fragments could be mutually reached if the GP map is
many-to-many. The existence of ``promiscuous'' sequences that map into more than 
one phenotype enhances navigability and promotes fast 
adaptation~\cite{bloom:2007b,catalan:2017}.

The statistical property of GP maps that has attracted the most attention is very
likely the distribution of genotype network sizes, or phenotype sizes for short. 
Due to the astronomically large sizes of genotype spaces, initial estimations of 
the size of phenotypes were performed through random samplings of genotype space.  
The results were often represented as frequency-rank plots, with phenotypes 
ordered according to their sizes. Random samplings of genotype spaces in 
many-to-one GP maps invariably yielded some very abundant phenotypes and a large 
number of phenotypes represented by a few or just one 
genotype~\cite{schuster:1994,gruner:1996}. Often, a frequency-rank plot was fitted
to a generalized Zipf's law~\cite{tacker:1996}, implying a power-law-like
distribution of phenotype sizes. However, subsequent studies demonstrated that
the frequency-rank plot of phenotype sizes actually had a more complex
functional shape~\cite{stich:2008,cowperthwaite:2008,irback:2002,catalan:2014},
and specific functional fits were avoided. Subsequent studies have exhaustively
mapped the complete sequence space to its corresponding phenotypes, among which
RNA sequence-to-minimum energy secondary structure 
map~\cite{cowperthwaite:2008,dingle:2015}, the hydrophobic-polar (HP) model for 
protein folding~\cite{irback:2002,holzgrafe:2011}, or toyLIFE, which includes a
sequence-to-structure-to-function description~\cite{catalan:2014}. As a result,
complete phenotype size distributions (for short sequences) are now available.
Fitted shapes range from power-law-like curves~\cite{ferrada:2012} to lognormal 
distributions~\cite{dingle:2015}.

It has been argued that, among other generic properties, a skewed distribution
of phenotype sizes results from the organization of biological sequences into
constrained and unconstrained parts. In~\cite{greenbury:2015}, the authors
introduce the Fibonacci GP map, a many-to-one artificial model, where sites in a 
sequence can be coding or non-coding, and either lead to new phenotypes under 
mutations (coding sites) or yield the same phenotype (neutral, non-coding sites). 
The model can be analytically solved and yields a power-law phenotype size 
distribution, in qualitative agreement with some observations. 

In this contribution, we attempt an identification of the elements in the
organization of sequences that characterize the {\it quantitative} properties of 
the distribution of phenotype sizes. We show in a constructive fashion that the 
model in~\cite{greenbury:2015} is an example of a broad spectrum of 
sequence-to-structure GP models. Starting with the simplest case, where sequences 
are separated into constrained and neutral parts, and adding subsequent elements in 
the organization of the sequences and versatility levels of the sites, we show how 
the distribution of phenotype sizes changes from pure power-law (with an exponent 
dependent on how genotypes are distributed among phenotypes) to lognormal. This 
functional form is independent of whether the GP map is many-to-many (sequences are 
promiscuous) or many-to-one (the phenotype can be uniquely predicted from the 
sequence). Our final example corresponds to the RNA sequence-to-secondary structure 
map, where we demonstrate that the combinatorial properties of the distribution of 
sites of variable neutrality along sequences causes the distribution of phenotypes 
to follow a lognormal distribution, with parameters that can be traced to 
properties of the genotype set. Our main result is that a lognormal distribution of 
phenotype sizes is the expected result in any GP map where sufficient variation in 
the number of phenotypes of similar size is present. 

\section{Definitions}
\label{sec:definitions}

\begin{figure}[h]
\centering
\includegraphics[width=\textwidth]{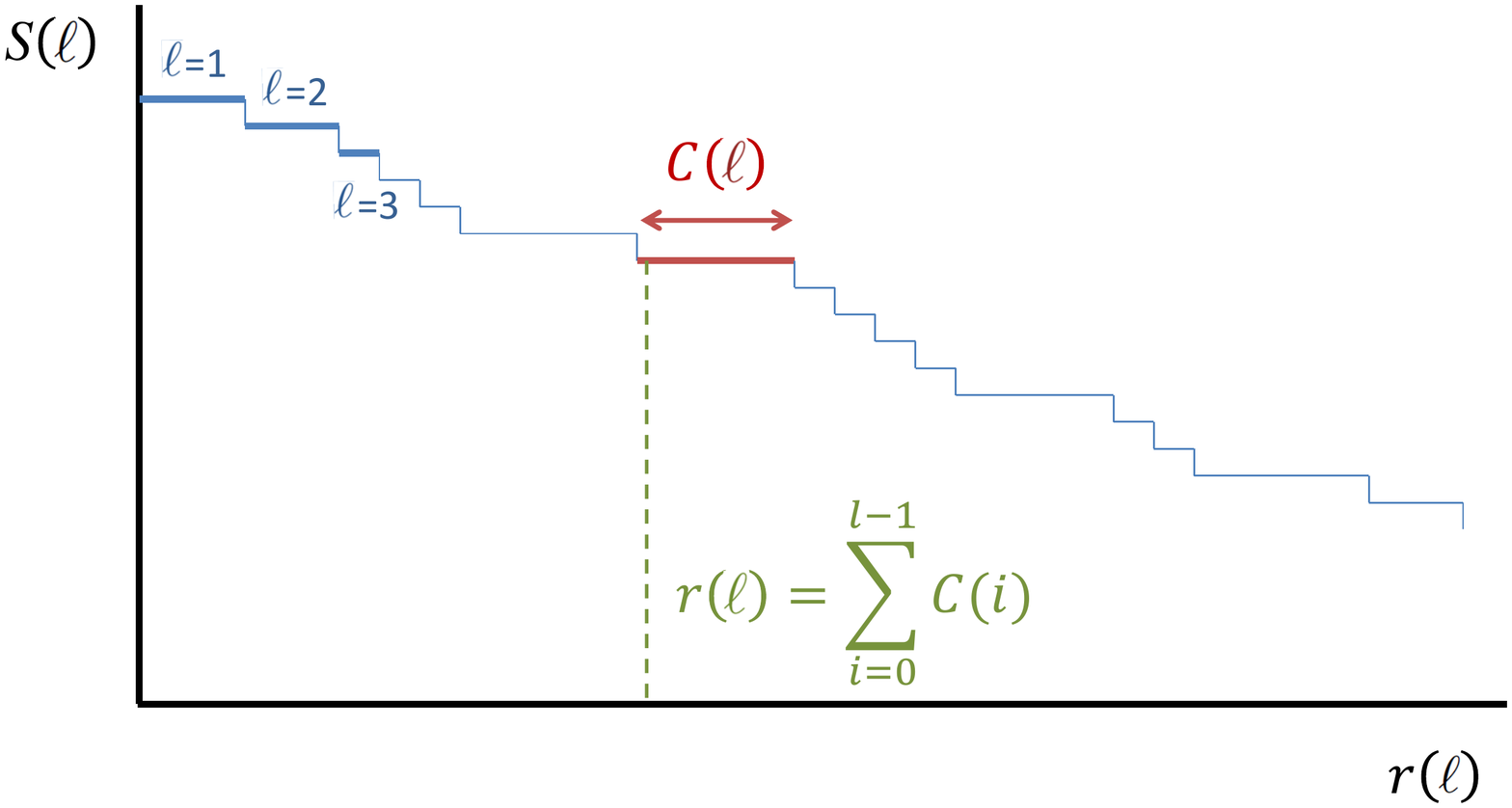}
\vspace*{-15mm}
\caption[]{Schematic of the quantities involved in the calculation of the abundance 
$S(r)$ of a phenotype as a function of its rank $r$. Here $\ell$ represents the 
number of constrained sites and works as an intermediate variable to simplify 
calculations; the first three values of $\ell$ are indicated in the figure. 
$C(\ell)$, corresponding to the length of the horizontal segments (arbitrary in this
representation), is the number of phenotypes with $\ell$ constrained sites and 
$S(\ell)$ is their size. 
}
\label{fig:quantities}
\end{figure}

We will study four models that interpolate between the simplest case of sequences 
divided into neutral and non-neutral sites separated into two groups and a general 
case (represented by RNA), and calculate for each of them the size of a phenotype 
given the sequence organization of its corresponding genotypes, the number of 
phenotypes with the same size, the frequency rank ordering of phenotypes, and 
eventually the distribution of phenotype sizes. Table~\ref{table:definitions} 
summarizes the nomenclature and definitions used in this work, and 
Figure~\ref{fig:quantities} illustrates some relevant quantities.

\begin{table}
\centering
\begin{tabular}{|c|l|}
\hline
Symbol & Definition \\
\hline
$L$ & Sequence or genotype length \\
$k$ & Alphabet size  \\
$v(i)$ & Versatility of site $i$  \\
$\ell$ & Number of sites in the low-versatility
class $v_2$\\
$L-\ell$ & Number of sites in the high-versatility class $v_1$\\
$S(\ell)$ & Size of a phenotype  \\
$C(\ell)$ & Number of phenotypes with the same size  \\
$N_c(\ell)$ & Set of $\ell$-genotypes \\
$r(\ell)$ & Rank of a phenotype  \\
$p(S)$ & Probability density that a phenotype has size $S$ \\
$Q(L,\ell)$ & Number of phenotypes from different ordering of sites \\
\hline
\end{tabular}
\caption{Summary of symbols used in this work and their short definitions.}
\label{table:definitions}
\end{table}

The genotype space is made of sequences of length $L$ letters from an alphabet
of size $k$. Two examples of alphabet sizes are $k=2$ for a binary alphabet
$\{0,1\}$ and $k=4$ for DNA or RNA, $\{${\tt{A, C, G, T}} or {\tt{U}}$\}$. The
\emph{versatility} $v(i)$ of site $i$ is defined as the average number of
different letters of the alphabet that can occupy a given sequence position
$i$. In general $k \ge v(i) \ge 1$ for all sites $i$. This is a quantity
closely related to neutrality. We will study the simplified case where sites
can take one out of two different values, $v(i) \in \{v_1,v_2\}$, with $k \ge
v_1 > v_2 \ge 1$. Sites are called constrained if $v_2=1$, and neutral if
$v_1=k$. We will use $\ell$ to count the number of sites with low versatility.

The size $S(\ell)$ of a phenotype is the number of different genotypes compatible 
with that phenotype. From the definition of $\ell$ it follows that $S(\ell)$ is
a nonincreasing function of $\ell$. In the literature, phenotype
frequency~\cite{greenbury:2015}, number of sequences for a
phenotype~\cite{ferrada:2012} or neutral set size~\cite{dingle:2015} have been
used with a meaning identical to phenotype size here. The set of
$\ell$-genotypes is defined as the number of genotypes compatible with
$\ell$-phenotypes, $N_c(\ell) \equiv S(\ell) C(\ell)$. The rank of the first
phenotype in size class $C(\ell)$ is $r(\ell)=\sum_{i=0}^{\ell-1} C(i)$. Note
that the total number of phenotypes coincides with the maximum rank.

If $p(S)$ is the probability density that a phenotype has size $S$, then we can
count phenotypes as
\[
C(\ell)=\sum_{i\ge\ell}C(i)-\sum_{i\ge\ell+1}C(i)=\Pr\{S\le S(\ell)\}-
\Pr\{S\le S(\ell+1)\}=\int_{S(\ell+1)}^{S(\ell)}p(S)\,dS.
\]
To first order we can approximate
the integral as $C(\ell)\approx p(S)|S'(\ell)|$ (the approximation gets better
the smaller $S'(\ell)$). Thus, up to a normalisation constant,
$p(S)\propto C\big(\ell(S)\big)\left|S'\big(\ell(S)\big)\right|^{-1}$.

The probability density $p(S)$ yields the probability of finding a phenotype with 
size $S$ when uniformly sampling over phenotypes. This corresponds to the 
distribution $P_P(S)$, as defined in other studies~\cite{dingle:2015}. 

Finally, we will also introduce a factor $w(\ell)$ to represent the fraction of
$\ell$-genotypes that actually go to a given $\ell$-phenotype. This factor arises
from additional restrictions in the assignment of genotypes to phenotypes which are
not made explicit in the models. In general, if $w(\ell)=1$ the models we are
going to introduce assign the same genotypes to several $\ell$-phenotypes.
This would correspond to a many-to-many GP map ---a sort of maps suitable to
describe molecular promiscuity. 
Incidentally, molecular promiscuity strongly enhances navigability in genotype 
space~\cite{bloom:2007b,piatigorsky:2007,catalan:2017}. Other
choices may account for specific restrictions in the models; in particular, a
suitable choice of $w(\ell)$ may render the GP map many-to-one. We will return
to this point when we provide details of the models.

A succint definition of the hierarchy of models introduced in this work is as
follows:

\begin{itemize}
\item {\bf Model 1:} {\em Constrained and neutral sites occupy fixed positions.} 
Sequences are separated in two parts, the first one of length $\ell$ occupied by
constrained sites, $v_2=1$, and the second part of length $L-\ell$ occupied by 
neutral sites, $v_1=k$. Two minor variants considered are (i) phenotypes are all 
viable and (ii) lethal mutations occur independently of the site class. 
\item {\bf Model 2:} {\em Constrained and neutral sites occupy variable positions.} 
This is illustrated by means of two examples: (i) constrained sites are split into 
two fragments at the beginning and at the end of the sequence and (ii) constrained 
sites can occupy arbitrary positions in the sequence.
\item {\bf Model 3:} {\em Versatile sites occupy fixed positions.} Two different 
types of sites with fixed versatilities $v_1$ and $v_2$ are considered.
\item {\bf Model 4:} {\em Versatile sites occupy variable positions: RNA.} In a
first approximation, RNA sequences contain two types of sites that occupy
different positions in the sequence subject to secondary structure constraints:
those forming pairs (stacks) in the secondary structure have average versatility 
$v_2$, and those unpaired (loops) have average versatility $v_1$. The model can be 
generalized to an arbitrary number of site classes.
\end{itemize}

\begin{figure}[t!]
\centering
\includegraphics[width=0.85\textwidth]{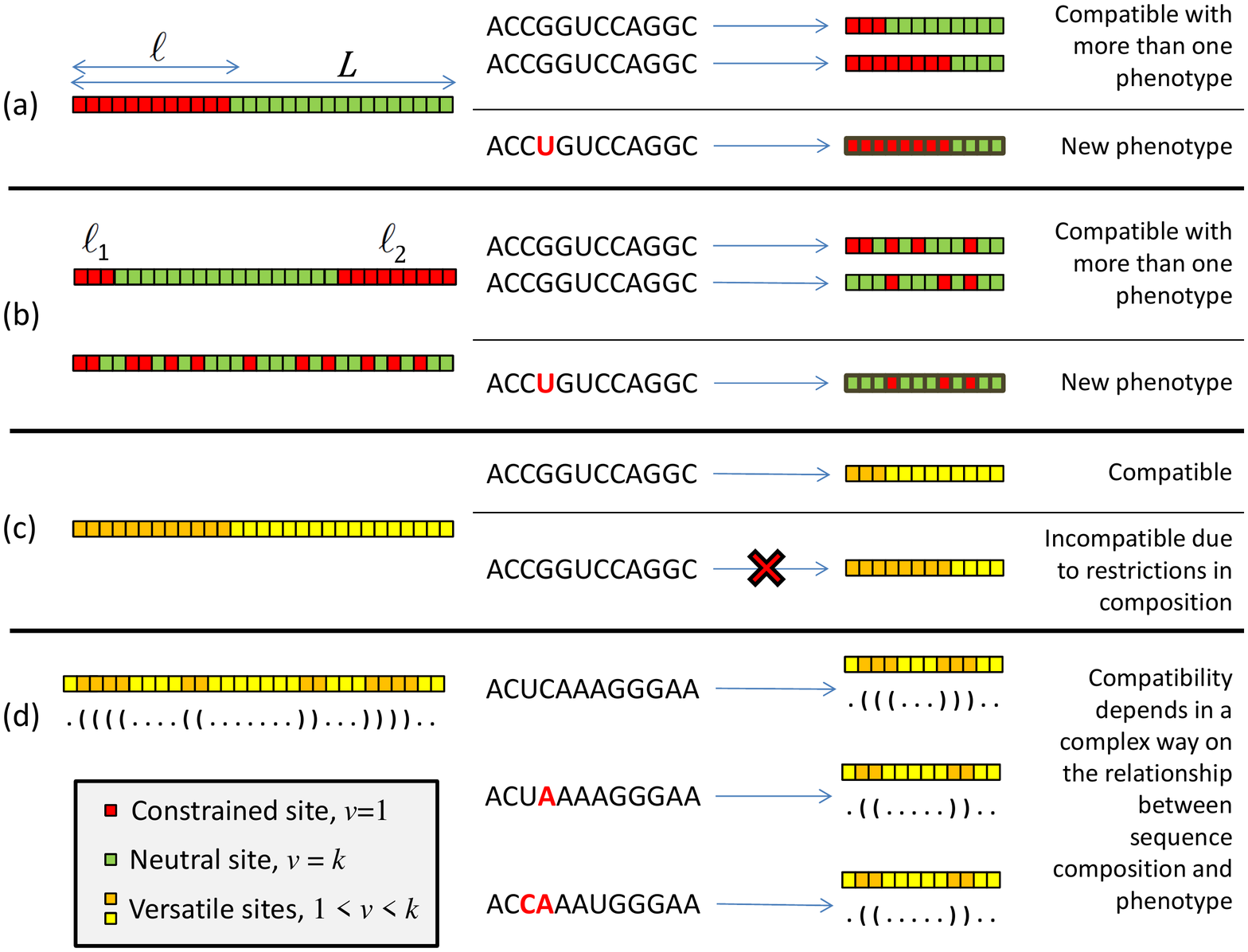}
\caption[]{\footnotesize Pictorial representation of the main models analysed
in this work. Each number represents a main class of models, as described in
Section~\ref{sec:definitions}. From
left to right, we depict the organization of sites in phenotypes, examples of how sequences 
are assigned to phenotypes --with example mutations highlighted in bold red, and a brief 
explanation on possible constraints in that assignment. 
(a) In Model 1 a given sequence can be assigned with equal probability to any phenotype 
with $\ell$ constrained sites. A mutation in one of those sites changes the 
phenotype, and there is only one possible structural arrangement given $\ell$. The bold
line in the phenotype associated to the mutated sequence is meant to represent a nwe phenotype
with the same arrangement as above.
(b) Two examples analysed in this work in the class of Model 2 are when constrained sites 
are split into two parts at the beginning and at the end of the sequence (above) and 
when they can occupy arbitrary positions (below). Different ways in which the $\ell$ 
constrained sites can be arranged define different phenotypes to which the same sequence
can be assigned. As in Model 1, a mutation in a constrained site changes to a new phenotype 
with the same structural arrangement. Models 1 and 2 are unconstrained, many-to-many maps.
(c) As it occurs in Model 1, $\ell$ defines the structure of the phenotype in Model 3, 
though there might be constraints in the assigment of sequences. In the example, assuming 
that less versatile sites admit only two letters, for instance {\tt A} and {\tt C}, 
implies that the sequence shown cannot be assigned to any phenotype with letters 
{\tt G} or {\tt U} in less versatile sites (positions 4 to 6 in the example). 
(d) Model 4 includes elements of Models 2 and 3: the order of sites with different 
versatilities matters in the definition of phenotype and there are restrictions in the 
assignment of sequences to phenotypes. In the example shown, corresponding to RNA, 
mutations may or may not change the phenotype, depending on a non-trivial relationship 
between structure and sequence composition. Still, in general, also in Models 3 and 4 a 
genotype can be assigned to multiple phenotypes.}
\label{fig:models}
\end{figure}

Figure~\ref{fig:models} schematically represents the different models analysed
here and some properties that will be of relevance to understand the distributions
of phenotype sizes they yield. 

\section{Results}

\subsection{Model 1: Constrained and neutral sites occupy fixed positions}

This is probably the simplest non-trivial model in the class of GP maps, very
similar in spirit to that presented in~\cite{greenbury:2015}. Phenotypes are
characterized by $\ell$ constrained sites in the first part of the sequence.
For a fixed $\ell$, mutations in a constrained site change the phenotype, and
mutations in neutral sites yield genotypes compatible with the phenotype. 
Therefore, 
\begin{eqnarray}
\label{Eq:fl}
S(\ell)&=&k^{L-\ell} w(\ell) \\
\label{Eq:Cl}
C(\ell)&=&k^{\ell} \\
\label{Eq:rl}
r(\ell)&=&\frac{k^{\ell}-1}{k-1}
\end{eqnarray}

Note that, if $w(\ell)=1$, the complete genotype space is partitioned among
$\ell$-phenotypes for every value of $\ell$. This implies that, if we consider
all possible phenotypes (i.e. all $\ell$ values), a particular genotype is
simultaneously compatible with many different phenotypes ---representing a highly 
promiscuous sequence. Specifically, if $w(\ell)=1$ the total number of genotypes
compatible with $\ell$-phenotypes is $N_c(\ell) = k^L$, so the total amount of
genotypes $\sum_{\ell} N_c(\ell)=(L+1)k^L$. This result clearly shows the
many-to-many nature of the GP map of this model with this choice of $w(\ell)$
---genotypes are assigned to all phenotypes they are compatible with and,
therefore, are repeatedly counted.

A minimal rule to avoid multiple assignments is to think of $w(\ell)$ as the
probability that a genotype is actually assigned to an $\ell-$phenotype. When
this probability is uniform, $w(\ell)=\Omega$, and we choose
$\Omega=(L+1)^{-1}$, the total number of genotypes becomes $\sum_{\ell}
N_c(\ell)=k^L$, the size of the genotype space, so the resulting map is effectively
many-to-one. Other examples in which $w(\ell)$ depends on $\ell$ will appear later. 

Now, to obtain size as a function of rank we must eliminate $\ell$ in $r(\ell)$
and substite it into $S(\ell)$ to get $S(r)$. In this case, from
Eq.~\eqref{Eq:rl} and assuming $(k-1) r \gg 1$, 
\begin{equation}  
\ell=\log_k\left[(k-1)r+1\right] \approx \log_k\left[(k-1)r\right],
\end{equation}
and substituting in~\eqref{Eq:fl}
\begin{equation}
S(r) \approx \Omega \frac{k^L}{k-1} r^{-1}.
\end{equation}

To obtain the probability density $p(S)$ we first notice that Eq.~\eqref{Eq:fl}
implies $k^{\ell}= \Omega k^L S^{-1}$, hence $C(S)= \Omega k^L S^{-1}$. On the other 
hand $S'(\ell)=-(\log k)S$, thus
\begin{equation}
p(S) \propto S^{-2}.
\end{equation}
Hence the probability distribution is a power-law with exponent $\beta=2$.

\subsubsection{Non-viable genotypes arise from uniformly distributed lethal
mutations}
\label{sec:lethal}

In the same scenario as above, let us assume that a fraction $\delta$ of mutations 
is lethal, thus leading to a non-viable genotype. In this case, Eqs.~\eqref{Eq:fl} 
to~\eqref{Eq:rl} are identical, with $k$ substituted by $k(1-\delta)$. Therefore, 
$S(r)$ and $p(S)$ are as above with the latter change. This result shows that the 
existence of a non-viable class to which viable genotypes can mutate does not 
necessarily imply relevant functional changes in the distribution of phenotypes, 
which is in either case of the form $p(S) \sim S^{-\beta}$, with $\beta = 2$. The 
effect of uniformly distributed lethal mutations could be therefore absorbed as a 
constant into $\Omega$. The situation changes if mutations are not distributed 
uniformly, but their likelihood depends on $\ell$. This would be a particular 
realisation of Model 3 introduced below.  

\subsection{Model 2: Constrained and neutral sites occupy variable positions}

In any realistic model (e.g. the case of RNA) the position of constrained and 
neutral sites should matter in the definition of a phenotype. While $S(\ell)$ does 
not change its functional form as a result, $C(\ell)$ does (and $r(\ell)$ as a 
consequence), causing potentially relevant modifications in $S(r)$ and $p(S)$. In 
general, the number of different phenotypes would take the form $C(\ell) = k^{\ell} 
Q(L,\ell)$, where $k^{\ell}$ accounts for changes in the letter of the constrained 
site (yielding a different phenotype, as assumed) and $Q(L,\ell)$ is a 
model-dependent combinatorial number that counts the different ways in which the 
$\ell$ sites can be arranged to yield meaningful (and different) phenotypes. In 
general, the factor $S^{-2}$ in $p(S)$ stems from mutations in neutral sites, while 
the arrangement of constrained and neutral sites along the sequence is weighted by 
$Q\big(L,\ell(S)\big)$, with effects on the functional form of $p(S)$ that, in 
general, depend on the permitted arrangements. As will be shown, $Q(L,\ell)$ might 
enormously increase the number of phenotypes and, especially, the relative 
abundances of $\ell$-phenotypes. 

\subsubsection{Constrained sites are split into two groups at the extremes of
the sequence}
\label{sec:model2a}

As a way of example, let us consider one of the simplest situations where the
position of the constrained sites matters. Suppose that those sites can be
split into two groups with lengths $\ell_1$ and $\ell_2$ and placed at the
beginning and at the end of the sequence (such that $0 \le \ell_1, \ell_2 \le
L$ and $\ell_1+\ell_2 = \ell$). This gives $Q(L,\ell)=\ell+1$ different
phenotypes with $\ell$ constrained sites, and
\begin{eqnarray}
S(\ell)&=&k^{L-\ell}\Omega, \label{eq:Sellmodel2} \\
C(\ell)&=& k^{\ell} (\ell+1), \label{eq:Cellmodel2} \\
r(\ell)&=& \frac{k^{\ell}(\ell k-\ell -1)+1}{(k-1)^2}.
\label{eq:rellmodel2}
\end{eqnarray}
From these expressions we can obtain (see Appendix~A) the asymptotic (for large
$r$) rank distribution
\begin{equation}
S(r)\propto\frac{\log r+a}{r},
\label{eq:rankmodel2}
\end{equation}
and the size probability density
\begin{equation}
\label{eq:pSmodel2}
p(S)\propto\frac{\log S+b}{S^2},
\end{equation}
with $a$ and $b$ some constants.

Therefore, even in this simple case with quite a limited number of possible
organization of constrained sites, $S(r)$ and $p(S)$ are no longer pure power-laws, 
though the dominant term of the phenotype size distribution (size still dominated 
by mutations in neutral sites) is characterized by an exponent $\beta = 2$. The 
total number of genotypes compatible with $\ell$-phenotypes is also modified, 
$N_c(\ell) = k^L (\ell+1)$, and is seen to increase linearly with $\ell$. 

\subsubsection{Constrained sites can occupy any position in the sequence}

We now assume that the constrained and unconstrained sites can occupy any site
of the chain. In that case
\begin{eqnarray}
S(\ell)&=&k^{L-\ell} \Omega,  \\
C(\ell)&=& k^{\ell} \binom{L}{\ell}  \, ,
\end{eqnarray}
with no simple expression for $r(\ell)$. Let us focus, however, on
the size distribution $p(S)$, and consider the case where $L\gg 1$.
Asymptotically for $L\to\infty$
\begin{equation}
\binom{L}{\ell}\sim 2^L\sqrt{\frac{2}{\pi L}}\exp\left\{-\frac{2}{L}
\left(\ell-\frac{L}{2}\right)^2\right\}.
\end{equation}
Changing $\ell$ to $S$ through $\ell=L+\log_k\Omega-\log_kS$
\begin{equation}
p(S)\propto\frac{1}{S^2}
\exp\left\{-\frac{2}{L}
\left(\log_kS-\frac{L}{2}-\log_k\Omega\right)^2\right\},
\end{equation}
and writing $S^{-1}=\exp(-\log k\,\log_k S)$, we finally obtain
\begin{equation}
p(S) \sim \frac{1}{S\log k}
\sqrt{\frac{2}{\pi L}}\exp\left\{-\frac{2}{L}
\left[\log_kS-\frac{L}{2}\left(1-\frac{\log k}{2}\right)
-\log_k\Omega\right]^2\right\}
\end{equation}
a log-normal distribution with mean $\mu_L\sim\frac{\log k}{2}
\left(1-\frac{\log k}{2}\right)L+\log\Omega$ and variance $\sigma_L^2\sim
\left(\frac{\log k}{2}\right)^2L$, very different from the $p(S)\sim S^{-2}$
distribution of the previous cases.

This section presents an example of a main result of this study. It shows that, 
when the definition of the phenotype depends on the specific position of 
constrained and neutral sites in sequences, the functional form of $p(S)$ (and, in 
consequence, of $S(r)$) qualitatively changes. In particular, the exponential 
growth of $Q(L,\ell)$ with $L$ dominates $p(S)$, which takes the form of a 
lognormal distribution. Other quantities defining the GP map, such as $k$ or 
$\Omega$, change now the parameters of the distribution, but do not modify its 
shape. 

\subsection{Model 3: Versatile sites occupy fixed positions}

The models analysed above demonstrate that when sites are either constrained or
neutral, the exponent associated to the power-law part of $p(S)$ is $\beta=2$. As 
we show next, this exponent is modified when the sites in the sequence show 
intermediate degrees of versatility, which causes the number of $\ell$-genotypes to 
depend on $\ell$.

Let us consider the case where the $L-\ell$ sites are just less constrained
than the $\ell$ sites, such that the former admit an average of $v_1$ different
letters of the alphabet and the latter admit $v_2$, with $k\ge v_1>v_2\ge 1$.
Relevant functions read
\begin{eqnarray}
S(\ell)&=&v_1^L\left( \frac{v_2}{v_1} \right)^{\ell} \Omega,
\label{eq:20} \\
C(\ell)&=& (k-v_1+1)^L \kappa^{\ell}, \\
r(\ell)&=& 
(k-v_1+1)^L \left( \frac{\kappa^{\ell}-1}{\kappa-1} \right),
\end{eqnarray}
with $\kappa \equiv (k-v_2+1)/(k-v_1+1)$. 

As it can be readily seen by substitution, these expressions reduce to Model 1
for $v_1=k$ and $v_2=1$. Now,
\begin{equation}
\ell = \log_{\kappa} \left(1 + \frac{\kappa-1}{(k-v_1+1)^L}r \right),
\end{equation}
yielding
\begin{equation}
S(r)= v_1^L \left( \frac{v_1}{v_2} \right)^{-\log_{\kappa}
\left(1 + \frac{\kappa-1}{(k-v_1+1)^L}r \right)}
=v_1^L\left(1 + \frac{\kappa-1}{(k-v_1+1)^L}r
\right)^{-\log_{\kappa}(v_1/v_2)}.
\end{equation}
For large $r$ this scales as $S(r)\sim c r^{-\alpha}$, where $\alpha$ depends
on $v_1$ and $v_2$ as
\begin{equation}
\alpha=\log_{\kappa}\left(\frac{v_1}{v_2}\right),
\end{equation}
yielding $\alpha=1$ in the limit of Model 1. Substituting this expression into 
Eq.~\eqref{eq:20},
\begin{equation}
\ell=-\frac{1}{\alpha}\log_{\kappa}\left(\frac{S}{v_1^L\Omega}\right),
\end{equation}
hence, up to a constant factor,
\begin{equation}
p(S) \propto \kappa^{-\frac{1}{\alpha}\log_{\kappa}S} S^{-1}\propto S^{-1-\alpha^{-1}}.
\end{equation}
Again $p(S)$ maintains its power-law shape but its exponent depends on $v_1$
and $v_2$.

The number of $\ell$-genotypes now becomes 
\begin{equation}
N_c(\ell)=\Omega v_1^L(k-v_1-1)^L\left(\frac{v_2}{v_1}\kappa\right)^{\ell}.
\end{equation}
This number can either increase or decrease with $\ell$ depending on whether
$v_2/v_1 \kappa$ is larger or smaller than 1. Both situations are possible
under the constraint $v_1 > v_2$. The values of $\alpha$ and $\beta$ change in
response to possible enrichements or depletions in the total number of assigned 
genotypes with $\ell$. This is a first example of similar cases encountered later
in this work and in the literature, as we discuss later. 

\subsection{Model 4: Versatile sites occupy variable positions: RNA}

In a first approximation (which has been shown to yield acceptable fits to
data~\cite{aguirre:2011}), RNA sequences can be divided into two classes of sites: 
those in stacks (bound) and those in loops (unbound), characterized by different 
degrees of neutrality (see e.g.~\cite{huynen:1996b} and Fig.~4 
in~\cite{reidys:2001}). Changes in the position of loops and stacks means a 
different phenotype. Additionally, the composition of each site in the sequence 
bears a significant correlation with the structural element it will preferentially 
represent in the phenotype (see Fig.~7 in~\cite{schultes:1997}). Therefore, a first 
approximation to a GP representation of RNA involves elements in our previous 
Models 2 and 3. In the following, the abundances or phenotypes will be ruled by 
(averaged) values $v_2$ and $v_1$ of the number of letters that can be changed in 
stacks or loops, respectively (see Fig.~\ref{fig:models}), without affecting the 
phenotype. 

Studies of RNA neutral networks and their related properties are usually restricted
to the many-to-one mapping between sequence and structure. Despite the fact that any
RNA sequence is compatible with multiple structures whose relative weight in an 
ensemble of identical sequences is defined by their folding 
energy~\cite{mccaskill:1990}, it is common practice to select only the minimum 
energy fold as the associated phenotype. This decision transforms an intrinsic 
many-to-many GP map where alternative phenotypes can be reached through mutations
or promiscuity, into a many-to-one map where navigability is limited to the effects 
of neutral drift. Analytical approaches cannot include, in general, 
energetic considerations, so they implicitly work in the many-to-many unrestricted 
case. This situation is comparable to the assignation of sequences to structures we 
have performed in our models, where every sequence is assigned to all phenotypes it 
is compatible with, while possible restrictions in the assignments are encompassed 
in $w(\ell)$. The distribution of secondary structure sizes for the unrestricted 
map (i.e. all sequences compatible with a given secondary structure) fixing the 
number of stacks or loops has been derived in~\cite{reidys:2002b} for the general 
case of structures with pseudoknots, in~\cite{poznanovic:2014} 
and~\cite{nebel:2002}, and in~\cite{cuesta:2017} in a form that will be used here. 

\subsubsection{Number of secondary structures with fixed number of pairs in
RNA}

In this case $\ell$ will denote the number of pairs of nucleotides in stacks
($\ell=1, 2,\dots, (L-j)/2$, with $j=3$ if $L$ is odd and $j=4$ if $L$ is
even), hence $L-2\ell$ will be the 
number of nucleotides in loops ($L-2\ell \ge 3$, which is the size of the minimal
---hairpin--- loop); $p_{L,\ell}$ is the probability distribution for secondary 
structures with $2\ell$ paired nucleotides, for sequences of length $L$ (in the 
limit $L, \ell \to \infty$). It has been 
shown~\cite{reidys:2002b,poznanovic:2014,cuesta:2017} that this distribution
behaves as a normal distribution in $\ell$ with mean $\mu_L=\mu
L+\mu_0+O\left(L^{-1}\right)$ and standard deviation $\sigma_L=\sigma
L^{1/2}+\sigma_0 L^{-1/2} +O\left(L^{-3/2}\right)$. In the case that structures
with stems with less than two base pairs or loops with less than three unpaired
bases are forbidden ---accounting for minimal energetic constraints--- we obtain 
$\mu\approx 0.28647\dots$, $\mu_0\approx -1.36502\dots$, 
$\sigma\approx 0.25510\dots$, and $\sigma_0\approx -0.00713\dots$ Note that 
different constraints will lead to different values of these quantities, but 
otherwise will not change the fact that $p_{L,\ell}$ is a normal distribution.
Finally, the number $Q(L,\ell)$ of different phenotypes of a sequence of length $L$
with $2\ell$ paired bases is given, in the limit $L, \ell\to\infty$, by
\begin{equation}
Q(L,\ell) \sim\frac{1}{\sqrt{2\pi}\sigma_L}e^{-(\ell-\mu_L)^2/2\sigma_L^2} Q_L,
\label{eq:QLRNA}
\end{equation}
with $Q_L \sim 1.48 L^{-3/2} (1.85)^L$ (see
\cite{reidys:2002b,poznanovic:2014,nebel:2002,cuesta:2017}).

\subsubsection{Size distribution}

In the case that the unpaired sites admit $v_1$ different letters and the paired 
sites $v_2$ letters ($1\le v_2<v_1\le k$), the size of a phenotype is given by 
$S(\ell)=v_1^{L-2\ell}v_2^{2\ell}$. Here, we will consider that a phenotype is
formed by all sequences compatible with that phenotype, thus setting
$\Omega=1$. We have
\begin{equation}
\ell=\frac{L\log v_1-\log S}{2\log\left(\frac{v_1}{v_2}\right)}.
\label{eq:elllognormal}
\end{equation}
Denoting
\begin{equation}
\mu_S=L\log v_1-\mu_L, \qquad \sigma_S=2\log\left(\frac{v_1}{v_2}\right)
\sigma_L,
\end{equation}
and noting that $p_L(S)= Q(L,\ell)/Q_L$, substitution of \eqref{eq:elllognormal}
into \eqref{eq:QLRNA} yields the log-normal distribution
\begin{equation}
p_L(S)\sim\frac{1}{\sqrt{2\pi}\sigma_SS}e^{-(\log S-\mu_S)^2/2\sigma_S^2}.
\end{equation}

\subsubsection{Rank distribution}

In the same two-sites approximation
\begin{equation}
C(\ell)\sim(k-v_2+1)^{2\ell}(k-v_1+1)^{L-2\ell}Q(L,\ell).
\end{equation}
The functional form of the rank $r(\ell)$ is derived in Appendix~B. After some 
algebra we arrive at 
\begin{equation}
S(r) \sim v_1^{L(1-2a)}v_2^{2aL}
\exp\left\{\eta L\sqrt{1-\frac{\log r}{cL}}\right\},
\qquad \eta\equiv\sigma\sqrt{8c}\log\left(\frac{v_1}{v_2}\right),
\end{equation}
with constants $a$ and $c$ depending on parameters of the combinatorial factor
$Q(L,\ell)$, see Appendix~B. 

\section{Discussion}
\label{sec:discussion}

The functional shape of the distribution of phenotype sizes is strongly dependent 
on the sequence organization within phenotypes. In a first approximation that 
discards the heterogeneity among genotypes in the same phenotype, one may describe 
that ensemble of sequences through a prototypic sequence whose sites admit a 
phenotype-dependent, variable number of letters of the alphabet, a quantity that we 
have dubbed versatility. The substitution of each sequence in a phenotype by the 
average over the phenotype seems a strong approximation. However, there is evidence 
that deviations from the average within a phenotype are small: the number of 
neutral neighbours of genotypes within a phenotype are tightly clustered around an 
average value characteristic of that phenotype size~\cite{aguirre:2011}. With this 
proviso, two main elements determine the corresponding distribution of phenotype 
sizes. The first one, generic for all systems, is the relationship between the size 
of a phenotype and the versatility $v(i)$ of each site $i$. In the framework used in
this work, the size of a phenotype can be written in general as 
\begin{equation} \label{eq:allometric}
S(\{v(i)\}) = \prod_i v(i).
\end{equation} 
This product yields an intrinsic allometric relation between the size of a 
phenotype and the length of the sequence. The second element, specific of each
sequence-to-structure map, is the number of phenotypes with similar size. This
quantity takes the overall form 
\begin{equation} 
\label{eq:comball}
C(\{v(i)\}) = Q(L, \{v(i)\}) \prod_i (k-v(i)+1),
\end{equation}
with the combinatorial factor accounting for the number of ways in which an 
ensemble of $L$ sites with $v(i)$ values can be arranged into meaningful 
phenotypes, and the product accounting for the number of neutral sequences within 
the phenotype. If the values of the combinatorial factor are constrained enough 
such that the asymptotic behavior of $Q(L, \{v(i)\})$ with $L$ is subdominant with 
respect to that of the product ---as in Models 1 and 3--- the distribution of 
phenotype sizes is a power-law. If, on the contrary, the dominant term is the 
combinatorial factor ---in particular when the distribution of structural motifs 
converges to a Gaussian--- the distribution of phenotype sizes becomes a lognormal.
Our calculations make it explicit that 
variations in the precise values of versatility, in the number of different classes 
of sites, or in particular constraints on structures (as, e.g. the minimum number 
of base pairs required to form a stack) have a quantitative effect on the 
parameters of the lognormal, but do not affect the shape of the distribution.

In the case $Q(L, \{v(i)\})\simeq 1$ we should expect a power-law-like distribution 
of phenotype sizes characterized by an exponent $\beta$. The actual value of 
$\beta$ stems from a combination of the number of genotypes compatible with a given 
phenotype and the total number of phenotypes with the same (or similar) size. 
Variations in the functional form of $w(\ell)$ with $\ell$ could be responsible for 
changes in $\beta$. In a general scenario, let us assume that phenotype sizes can 
be ordered according to a certain variable $\lambda$ (in our case the number of
low versatility positions $\ell$), and let us define the total 
number of genotypes compatible with $\lambda$-phenotypes as $N_c(\lambda) 
\equiv S(\lambda) C(\lambda)$, formally generalizing the quantity calculated in 
the specific models tackled in this work. The behaviour of 
$N_c(\lambda)$ with $\lambda$ determines the value of the exponent $\beta$:  If 
$N_c(\lambda)$ is constant, then $\beta=2$. However, if $N_c(\lambda)$ is 
exponentially enriched (depleted) in genotypes as $\lambda$ grows, the value of 
$\beta$ becomes larger (smaller) than 2. In the case of Model 3, for example 
$N_c(\ell)=A B^{\ell}$, with $B=(v_2/v_1) (k-v_2+1)/(k-v_1+1)$ and 
$\beta=1+1/\alpha$. Two examples of enrichment or depletion in the number of 
genotypes compatible with $\ell$-phenotypes are $\{v_1,v_2\}=\{4,2.5\}$, with 
$B=1.56$ and $\beta=2.95$, and $\{v_1,v_2\}=\{3,1.5\}$, with $B=0.875$ and 
$\beta=1.81$. In a very explicit way now, changes in the actual assignment of 
genotypes to phenotypes through $w(\lambda)$ (embedded in $S(\lambda)$) will 
affect the probability density distribution.

\begin{figure}[t!]
\centering
\includegraphics[width=\textwidth]{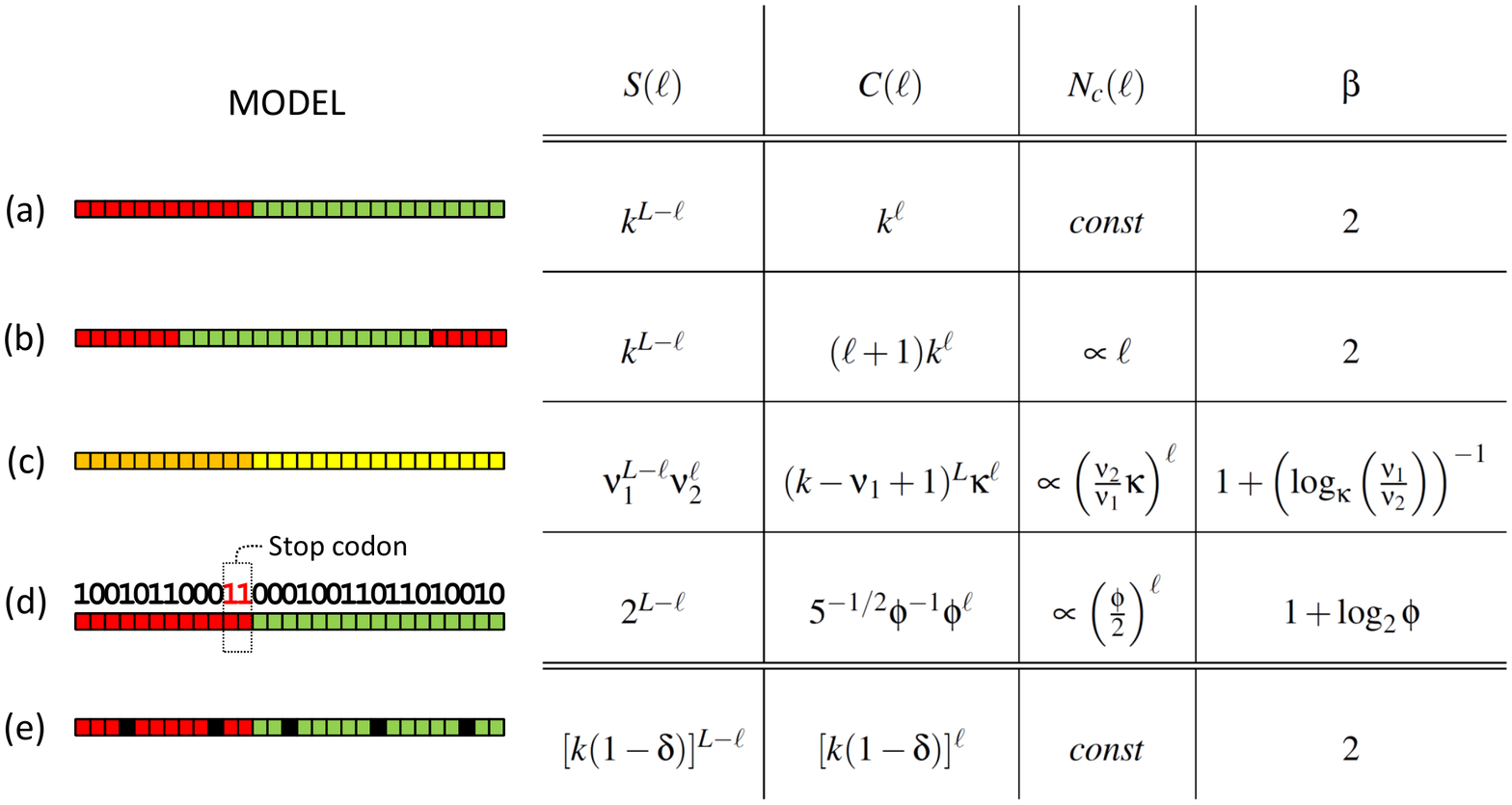}
\vspace*{-30mm}
\caption[]{Summary of constrained models yielding power-law-like distributions
of phenotype sizes and main analytical quantities. (a) Model 1: Constrained 
and neutral sites occupy fixed positions; (b) Model 2(i): Constrained sites are 
split into two groups at the extremes of the sequence; (c) Model 3: Versatile 
sites occupy fixed positions; (d) Fibonacci GP map~\cite{greenbury:2015} belongs to 
the class of model 3 and is analogous to models with an $\ell-$dependent rate of 
lethal mutations; (e) Model 1 with a uniform distribution of lethal mutations.
Colour codes for sites as in Fig.~\ref{fig:models}.}
\label{fig:powerlaw}
\end{figure}

Another example in the class of Model 3, yielding power-law-like $p(S)$ with 
non-trivial $\beta$ is the model in~\cite{greenbury:2015}. Besides the division of 
sequences into neutral and constrained sites, the authors introduce a stop codon 
which causes an $\ell-$dependent transition rate to alternative phenotypes, that 
being the eventual reason for a non-trivial value of $\beta$. In that case, 
$N_c(\ell)\approx 2^{L-\ell} \phi^{\ell-1}/\sqrt{5}$, which corresponds to a value of 
$B=0.81$ and, consistently, $2>\beta=1.69$, with $w(\ell)=\phi^{\ell-1}/(2^{\ell} 
\sqrt{5})$. The stop codon represents a particular instance of a decreased 
tolerance to mutations in less versatile sites. Another formal example could be
a rate to lethal mutations increasing with $\ell$. This class of mechanisms skew 
the assignation of genotypes to phenotypes or, equivalently, deplete the amount of 
genotypes associated to phenotypes as $\ell$ grows: larger values of $\ell$ imply 
that there are more positions where non-neutral mutations can occur, and this 
leads to $\alpha>1$ and $\beta<2$. Figure~\ref{fig:powerlaw} summarizes the 
sequence organization of different models with a power-law distribution of 
phenotype sizes, the origin and functional form of the $N_c(\ell)$ function, and 
the corresponding $\beta$ value.

\begin{figure}[t!]
\centering
\includegraphics[width=0.8\textwidth]{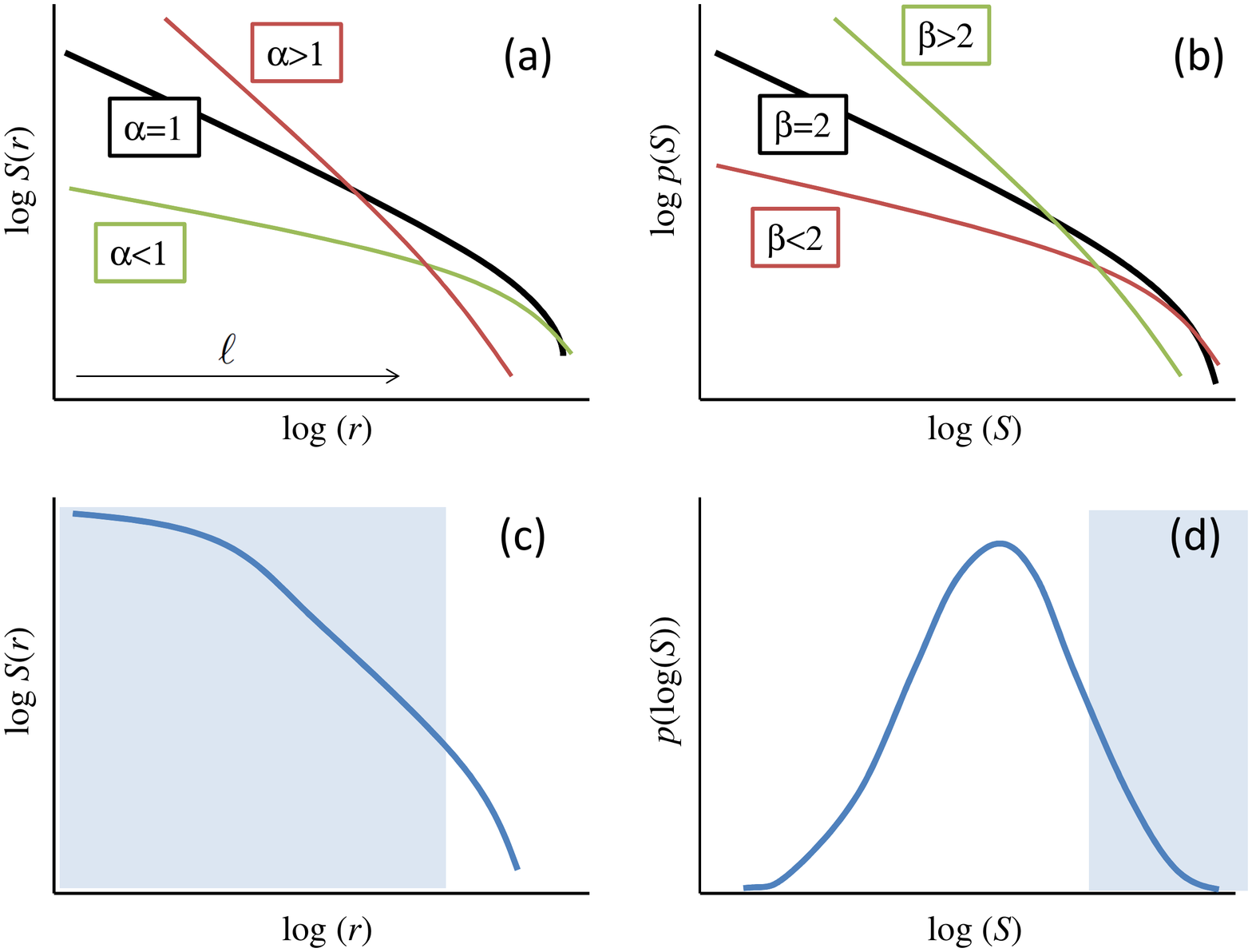}
\caption[]{Schematic representation of the functional form of $S(r)$ and $p(S)$ 
expected for different model classes. (a) Rank ordering of sizes in Model 1 yields 
$\alpha=1$. In the class represented by our model 3 (with additional examples in 
Figure~\ref{fig:powerlaw}) an enrichment or depletion of $N_c(\ell)$ with $\ell$ 
causes deviations towards smaller or larger values of $\alpha$, respectively, in 
absolute value. (b) Probability density $p(S)$: recall that $\beta=1+\alpha^{-1}$.
Colours code for equivalent curves in (a). (c) Possible shape of $S(r)$ for RNA 
(see, e.g.~\cite{stich:2008}) and expected $p(S)$ (d). Only sufficiently large 
phenotypes can be found under random sampling of genotypes (compare with results 
in~\cite{dingle:2015}); they would occupy the shaded blue regions in (c) and (d).}
\label{fig:distributions}
\end{figure}

In Fig.~\ref{fig:distributions} we represent schematically the functional form of 
$S(r)$ and $p(S)$ for the class of our Model 3 and a possibly general class of models 
analogous to RNA (class 4). At present, it is difficult to clearly match all models 
in the literature to classes 3 or 4. For example, the hydrophobic-polar (HP) non-compact 
model seems to be characterized by a distribution of phenotype sizes similar to a 
power-law~\cite{ferrada:2014}, while other models for heteropolymers that have been 
compared to HP yield broad distributions with a maximum~\cite{li:2002}. Even RNA with a 
two-letter alphabet apparently yields power-laws~\cite{ferrada:2012}, so it might belong 
to a non-trivial combination of models 3 and 4 as well. This is a very intriguing and 
complex question that we have to leave for future studies. These considerations 
notwithstanding, the situation where the combinatorial factor converges to a Gaussian 
distribution is expected to be very general for sequence-to-structure GP 
maps~\cite{poznanovic:2014}, implying that a lognormal distribution of phenotype sizes 
might be a generic property of such maps. Up to now, there are few quantitative results 
supporting this statement, very likely due to the impossibility to exhaustively fold 
genome spaces for large $L$. A remarkable exception is~\cite{dingle:2015}, where the 
lognormal distribution has been suggested as the best fit to computational distributions 
of RNA secondary structure sizes for lengths up to $L=126$. It is interesting to 
highlight that our results have been obtained under a uniform assignment of genotypes 
(represented through our variable $\Omega$) to phenotypes. However, the many-to-one GP 
map in RNA assigns the minimum energy structure to each sequence. In the language of 
our function $w(\ell)$, the correlation between energy and $\ell$ in RNA will 
preferentially assign genotypes to phenotypes with a large number of pairs (large 
$\ell$) since, on average, the larger the number of pairs the lower the folding 
energy~\cite{stich:2008}. It cannot be discarded that genotype-to-phenotype 
assignment rules based on quantities not considered here might skew the 
distribution or eventually yield different functional forms. Though this is a 
possibility that has to be kept in mind, results in~\cite{dingle:2015} reveal that, 
at least in the case of four-letters RNA, deviations from lognormality cannot be numerically 
detected. We suspect that this is likely due to a dominant effect of $Q(L,\ell)$ 
over $w(\ell)$ both in the many-to-many and in the many-to-one representations of 
the RNA sequence-to-structure map. 

Simple models as those presented here can be used as well to estimate other relevant
quantities of GP maps, and to determine if they are almost universal or 
model-dependent. One such quantity is the relationship between phenotypic 
robustness and the size of a phenotype. In our scenario, and similarly to other 
examples~\cite{greenbury:2015,greenbury:2016}, phenotypic robustness coincides with 
genotypic robustness, which is calculated straight forward as the ratio between the 
number of neutral neighbours, $(\nu_1-1)(L-\ell)+(\nu_2-1)\ell$ and the total 
number of neighbours of a sequence, $L(k-1)$. This yields a function of $\ell/L$. 
Next, $\ell$ is obtained easily from its relationship with $S(\ell)$, and it takes 
the general form $\ell \propto \log_{\zeta} S$, where $\zeta$ is a model-dependent 
quantity. Therefore, the relationship between phenotype robustness and the 
logarithm of phenotype size consistently appears in very generic 
sequence-to-structure models. The relationship between phenotype robustness and 
evolvability cannot be derived unless a explicit rule linking possible mutations to 
phenotypes with different $\ell$ is introduced. In our Models 1, 2, and 3, such a 
rule, which could take a form analogous to the stop codon of the Fibonacci 
map~\cite{greenbury:2015}, is not defined. The case of RNA is particularly 
interesting and has received significant computational attention since long 
ago~\cite{huynen:1996b}. Only partial explorations of the accessibility of 
alternative phenotypes have been performed due to the huge sizes of 
phenotypes~\cite{cowperthwaite:2008,schaper:2014}. Hopefully, further extensions of 
our Model 4 could help in the analytical treatment of this highly complex problem. 
Advances in empirical techniques, such as the intensive use of microarrays, should 
allow in the near future an exhaustive characterization of actual genotype spaces, 
as has been done for short transcription factor binding sites~\cite{payne:2014}. We 
believe that analyses of empirical GP maps will reveal strengths and weaknesses of 
the approach here presented, and likely suggest ways of improvement, regarding in 
particular a formal description of phenotype networks (networks of genotype
networks) and evolvability in natural systems.

\section*{Appendix A}

In order to derive $S(r)$ and $p(S)$ for Model~2 with constrained sites split
into two groups at the extremes of the sequence (Section~\ref{sec:model2a}) it
will prove convenient to use the affine transformation of $\ell$
\begin{equation}
x\equiv c_k\big[(k-1)\ell-1\big]=\ell\log k-c_k, \qquad
c_k\equiv\frac{\log k}{k-1}.
\end{equation}
Then Eq.~\eqref{eq:rellmodel2} can be rewritten
\begin{equation}
(k-1)^2r-1=\big[(k-1)\ell-1\big]k^{\ell}=\frac{x}{c_k}e^{\ell\log k}
=\frac{x}{c_k}e^{x+c_k},
\end{equation}
from which
\begin{equation}
xe^x=\left[(k-1)^2r-1\right]c_ke^{-c_k}.
\label{eq:productexp}
\end{equation}
Inversion of this equation yields
\begin{equation}
x=W\Big(\left[(k-1)^2r-1\right]c_ke^{-c_k}\Big),
\end{equation}
with $W(x)$ Lambert's product-logarithm function
\cite[Def.~4.13.1]{olver:2010}.

Now,
\begin{equation}
S(\ell)=\Omega k^Le^{-\ell\log k}=\Omega k^Le^{-c_k-x},
\end{equation}
and using Eq.~\eqref{eq:productexp},
\begin{equation}
S(\ell)=\frac{\Omega k^L}{c_k}\frac{x}{(k-1)^2r-1}.
\end{equation}
Finally, since $W(z)\sim\log z+O(\log\log z)$ when $z\gg 1$
\cite[Prop.~4.13.10]{olver:2010}, when the rank $r$ is large
\begin{equation}
x\sim\log\left[(k-1)^2r-1\right]+\log c_k-c_k\sim \log r+a,
\end{equation}
with $a$ a constant. Then, for large $r$ we obtain Eq.~\eqref{eq:rankmodel2}.

As for $p(S)$, from Eqs.~\eqref{eq:Sellmodel2} and \eqref{eq:Cellmodel2},
\begin{align}
C(\ell) &=(\ell+1)k^{\ell}=(\ell+1)\frac{\Omega k^L}{S}, \\
\log S &=\log\left(\Omega k^L\right)-\ell\log k.
\end{align}
Differentiating $\log S$ with respect to $\ell$ yields $|S'(\ell)|=S\log k$.
Therefore, eliminating $\ell$ from this same equation we end up with
\begin{equation}
C\big(\ell(S)\big)\left|S'\big(\ell(S)\big)\right|^{-1}
\propto\frac{\log S+b}{S^2},
\end{equation}
with $b$ another constant. This is Eq.~\eqref{eq:pSmodel2}.

\section*{Appendix B}

The rank function for the case of RNA sequences whose sites may take two values
of neutrality $v_1$ and $v_2$, a number $Q(L,\ell)$ of secondary structures of
length $L$ with $\ell$ sites with neutrality $v_1$ and a total number of $Q_L$
different secondary structures of lenght $L$ is
\begin{equation}
\begin{split}
r(\ell)\sim&\, Q_L (k-v_1+1)^L
\int_{-\infty}^{(\ell-\mu_L)/\sigma_L}\frac{1}{\sqrt{2\pi}}
\left(\frac{k-v_2+1}{k-v_1+1}\right)^{2\sigma_Lx+2\mu_L}e^{-x^2/2}\,dx \\
=&\, Q_L (k-v_1+1)^L\exp\left\{\mu_L\xi+\frac{\xi^2}{2}\sigma_L^2\right\}
\int_{-\infty}^{(\ell-\mu_L-\xi\sigma_L^2)/\sigma_L}\frac{1}{\sqrt{2\pi}}
e^{-x^2/2}\,dx,
\end{split}
\end{equation}
where
\begin{equation}
\xi\equiv2\log\left(\frac{k-v_2+1}{k-v_1+1}\right).
\end{equation}
Now, since $\ell-\mu_L-\xi\sigma_L^2$ will be negative for all
$\mu_L-\sigma_L\lesssim \ell\lesssim\mu_L+\sigma_L$, we can use the asymptotic
expansion of the complementary error function
\[
\erfc x\equiv\frac{2}{\sqrt{\pi}}\int_x^{\infty}e^{-t^2}\,dt=
\frac{2}{\sqrt{\pi}}\int_{-\infty}^{-x}e^{-t^2}\,dt\sim
\frac{e^{-x^2}}{x\sqrt{\pi}}
\]
to write
\begin{equation}
r(\ell)\sim 
\frac{Q_L \sigma_L(k-v_1+1)^L}{\sqrt{2\pi}(\mu_L+\xi\sigma_L^2-\ell)}
\exp\left\{\mu_L\xi+\frac{\xi^2\sigma_L^2}{2}-\frac{(\mu_L+\xi\sigma_L^2-\ell)^2}
{2\sigma_L^2}\right\}.
\label{eq:rankRNA}
\end{equation}

In order to find how the size of a phenotype depends on its rank value $r(\ell)$
it is convenient to introduce new parameters. Let us denote $\mu\equiv\mu_L/L$
and $\sigma\equiv\sigma_L/\sqrt{L}$, and
\begin{equation}
a\equiv\mu+\xi\sigma^2, \qquad
c\equiv\xi\mu+\frac{\xi^2\sigma^2}{2}+\log(k-v_1+1) +\log\rho
\end{equation}
with $\rho\simeq 1.85$. The size of a phenotype is given by
$S(\ell)=v_1^{L-2\ell}v_2^{2\ell}$, therefore
\begin{equation}
\frac{1}{L}\log S=\log v_1-2\frac{\ell}{L}\log\left(\frac{v_1}{v_2}\right).
\end{equation}
Now, taking logarithms in \eqref{eq:rankRNA} and neglecting subdominant terms
in $L$,
\begin{equation}
\frac{1}{L}\log r\sim c-\frac{1}{2\sigma^2}\left(a-\frac{\ell}{L}\right)^2.
\end{equation}
Hence
\begin{equation}
\frac{\ell}{L}\sim a-\sigma\sqrt{2c}\sqrt{1-\frac{\log r}{cL}}
\end{equation}
and therefore
\begin{equation}
\frac{1}{L}\log S\sim\log v_1-2a\log\left(\frac{v_1}{v_2}\right)
+\sigma\sqrt{8c}\log\left(\frac{v_1}{v_2}\right)
\sqrt{1-\frac{\log r}{cL}}
\end{equation}
which implies
\begin{equation}
S\sim v_1^{L(1-2a)}v_2^{2aL} \exp\left\{\eta L\sqrt{1-\frac{\log r}{cL}}
\right\}, \qquad \eta\equiv\sigma\sqrt{8c}\log\left(\frac{v_1}{v_2}\right).
\end{equation}

\section*{Author contributions}
SM and JAC designed the study, carried out the calculations, interpreted 
the results and wrote the manuscript. Both authors read and approved the final
text. 

\section*{Acknowledgements}
The authors acknowledge the thorough revision of three anonymous reviewers, which has
helped improving this paper.

\section*{Funding statement}
This work has been supported by the Spanish Ministerio de Econom\'{\i}a y 
Competitividad and FEDER funds of the EU through grants ViralESS (FIS2014-57686-P) 
and VARIANCE (FIS2015-64349-P).




\end{document}